\documentclass[twocolumn,prd,floatfix,amsmath,nofootinbib,amssymb,floatfix]{revtex4}
\usepackage{graphicx,color,dcolumn,booktabs,bm}
\usepackage{longtable,lscape}
\usepackage{txfonts}
\usepackage{overpic}
\usepackage{amssymb}
\usepackage{indentfirst}
\usepackage{feynmf}
\usepackage{slashed}
\usepackage{cases}
\usepackage{multirow}
\usepackage[colorlinks,
            citecolor=blue,
            anchorcolor=red,
            menucolor=red,
            linkcolor=red,
            filecolor=red,
            runcolor=red,
            urlcolor=blue,
            frenchlinks=red]{hyperref}
\usepackage{epstopdf}
\usepackage{bm}
\usepackage{makecell}
\usepackage{threeparttable}
\begin{document}
%******************************************************************
\title{Where are $3P$ and higher $P-$wave states in the charmonium family?}
\author{Ming-Xiao Duan$^{1,2}$}\email{duanmx16@lzu.edu.cn}
\author{Xiang Liu$^{1,2,3}$\footnote{Corresponding author}}\email{xiangliu@lzu.edu.cn}
\affiliation{
$^1$School of Physical Science and Technology, Lanzhou University, Lanzhou 730000, China\\
$^2$Research Center for Hadron and CSR Physics, Lanzhou University and Institute of Modern Physics of CAS, Lanzhou 730000, China\\
$^3$Lanzhou Center for Theoretical Physics, Key Laboratory of Theoretical Physics of Gansu Province and Frontiers Science Center for Rare Isotopes, Lanzhou University, Lanzhou 730000, China}

\begin{abstract}
How to hunt for higher $P$-wave states of charmonium is still an open topic when $2P$ charmonia were identified. {In this work, we present an unquenched quark model calculation to illustrate the spectroscopy behavior of these $3P$, $4P$ and $5P$ states in charmonium family.} For the $3P$ charmonia, the predicted masses are around 4.2 GeV and their two-body open-charm decay behaviors were given, by which we propose that searching for these $3P$ states via their open-charm decay channels from $\gamma\gamma$ fusion and $B$ decay can be accessible at future experiment like LHCb and Belle II.
We continue to calculate the masses of these $4P$ and $5P$ charmonia. Combing with these calculated results of  higher $P$-wave states of charmonium, we find that the coupled-channel effect becomes more obvious with increasing the radial quantum number, which can be understood well by the modified Godfrey-Isgur model with screened potential.
\end{abstract}
\pacs{11.55.Fv, 12.40.Yx ,14.40.Gx}
\maketitle

%******************************************************************
\section{Introduction}\label{sec1}
%******************************************************************

As the frontier of particle physics, how to quantitatively depict non-perturbative behavior of strong interaction has attracted extensive attention from both theorist and experimentalist. Studying hadron spectroscopy which has close relation to the dynamics of quark interaction can provide key hint to deepen our understanding of non-perturbative behavior of strong interaction. With the promotion of experimental precision, more and more charmoniumlike $XYZ$ states were observed in the past two decades.  {Obviously, it is a good chance to identify exotic hadronic states \cite{Liu:2013waa} and construct conventional charmonium family \cite{Brambilla:2010cs}} (see review articles \cite{Chen:2016qju, Liu:2019zoy,Guo:2017jvc} for recent progress).

Focusing on the charmonium family, we may notice that most of charmonia collected by the Particle Data Group (PDG) were found in the 1980s, which inspired the Cornell model \cite{Eichten:1974af}. Among these reported charmonia, vector charmonia is the main body, which include $J/\psi$ \cite{Aubert:1974js, Augustin:1974xw}, $\psi(3686)$ \cite{Abrams:1974yy}, $\psi(3770)$ \cite{Rapidis:1977cv}, $\psi(4040)$ \cite{Goldhaber:1977qn}, $\psi(4160)$ \cite{Brandelik:1978ei}, and $\psi(4415)$ \cite{Siegrist:1976br}. The exploration of vector charmonia with mass above 4 GeV may shed light on the $Y$ problem \cite{Wang:2019mhs, Wang:2020prx} involving in charmoniumlike $XYZ$ states (see Refs. \cite{Chen:2016qju, Liu:2019zoy} for more details). Additionally, most of the remaining discovered charmonia in the $J/\psi$ family correspond to $1P$ states, which are $h_c(3525)$, $\chi_{c0}(3414)$, $\chi_{c1}(3510)$, and $\chi_{c2}(3556)$. After finding out $X(3872)$ \cite{Choi:2003ue}, $X(3915)$ \cite{Uehara:2009tx}, and $Z(3930)$ \cite{Uehara:2005qd}, constructing $2P$ charmonium states becomes possible \cite{Duan:2020tsx, Liu:2009fe, Chen:2012wy, Chen:2013yxa}, which means that the radial quantum number of $P$-wave states accessible at the present experiment only reaches up to 2.
Facing on this situation, we may ask an interesting question: where are $3P$ and higher $P$-wave charmonia?

Until now, theorists have not paid enough attention to the studies on the $3P$ and higher $P$-wave states of charmonium due to the absence of experimental data.
For the $3P$ charmonia, the calculation from the quenched quark model like the Godfrey-Isgur (GI) model \cite{Godfrey:1985xj}
suggests that their masses are located in the range of $4.25\sim4.33$ GeV. Additionally, the GI model can also predict  the masses $4P$ states to be around 4.6 GeV \cite{Godfrey:1985xj}\footnote{{Although we give a short review of the study of charmonium, it is far from being complete. In the last decades of last century, there were bunches of researches on quarkonium spectroscopy, including, but not limited to, non-relativistic calculation \cite{Eichten:1994gt, Godfrey:1986wj, Buchmuller:1980su, Grant:1995hf, Fulcher:1991dm, Pantaleone:1985uf, Fulcher:1994ek, Barnes:1996ff}, Bethe-Salpeter equation \cite{Maris:1997hd, Maris:1999nt, Maris:1997tm, Frank:1995uk}, etc. The interesting reader may refer to the review article of \cite{Voloshin:2007dx, QuarkoniumWorkingGroup:2004kpm}. }}.
%{\color{red} In the nonrelativistic quenched potential model Ref.~\cite{Barnes:2005pb}, the masses and widths of $3P$ states were also discussed. Their results indicate the masses of $3P$ charmonia are from 4.20 to 4.32 GeV. Moreover, in Ref.~\cite{Deng:2016stx}, the branching fractions in the radiative decay processes of $3P$ states were calculated, by which the branching fractions were predicted in the $\mathcal{O}(10^{-3})$ order of magnitude. Additionally, the results of $3P$ charmonia are also shown in Refs.~\cite{Li:2009zu, Hao:2019fjg, Ono:1983rd, Zhou:2013ada, Badalian:1999fe}. }
There were some recent coupled-channel calculations of the higher $P-$wave charmonia in Refs.~\cite{Ferretti:2020civ, Ferretti:2018tco, Ferretti:2021xjl}. The authors of Ref.~\cite{Ferretti:2020civ} assigned the $X(4274)$ as $\chi_{c1}(3P)$ state, and further tried to categorize the reported charmoniumlike states $X(4500)$ and $X(4700)$ as the $\chi_{c0}(4P)$ and $\chi_{c0}(5P)$ state, respectively~\cite{Ferretti:2021xjl}. In fact,
we still need to make more efforts when establishing the $3P$ state and higher $P$-wave states of charmonium. Obviously, this opportunity is being left to us.

Before carrying out the present work, we have accumulated rich experiences when identifying $2P$ charmonium states \cite{Duan:2020tsx, Liu:2009fe, Chen:2012wy, Chen:2013yxa}. By adopting an unquenched quark model \cite{Duan:2020tsx}, we calculated the mass spectrum of the $\chi_{cJ}(2P)$ states and found that the mass gap between $\chi_{c0}(2P)$  and $\chi_{c2}(2P)$ can reach 13 MeV, which is consistent with the mass difference between $Z(3930)$ and $X(3915)$.
This observation also supports $X(3915)$ as $\chi_{c0}(2P)$ state \cite{Liu:2009fe}. Soon after, the LHCb Collaboration confirmed our prediction by analyzing the $B\to D\bar{D}K$ process, where $X(3915)$ as the $\chi_{c0}(2P)$ state was established definitely \cite{Aaij:2020ypa, LHCb:2020bls}.
Obviously, these established $2P$ charmonium states can be as good scaling point if further exploring higher states of $P$-wave charmonium.

Borrowing the research experience of establishing $2P$ charmonium states \cite{Duan:2020tsx, Liu:2009fe, Chen:2012wy, Chen:2013yxa}, we find that the coupled-channel effect should be seriously considered if obtaining the information of mass spectrum of $3P$ and higher $P$-wave charmonia, since more open-charm decay channels for these $3P$ and higher $P$-wave charmonia are allowed. In this work, we adopt an unquenched quark model \cite{Duan:2020tsx}, which was  once applied to successfully depict these $2P$ charmonia as given by experiment \cite{Aaij:2020ypa, LHCb:2020bls}. Along this line, we may continue to study these $4P$ charmonium states and even higher $P-$wave charmonia.
In our calculation, the mass spectrum and strong decay behavior of these discussed $3P$ and higher $P$-wave charmonia are given, which are valuable to further experimental searches for them.

This work is organized with three parts. After the Introduction in Sec.~\ref{sec1}, we illustrate the adopted unquenched quark model and give the corresponding numerical results of these
$3P$ and higher $P-$wave charmonia (see Sec. \ref{sec2}). Finally, this work ends with the summary in Sec.~\ref{sec3}.

\section{Predicting the $3P$ states and higher $P$-wave states of charmonium under an unquenched quark model}\label{sec2}
\subsection{The adopted unquenched quark model}\label{subsec21}

The Cornell model was constructed to depict the interaction between quarks and anti-quarks \cite{Eichten:1974af} with the accumulation of abundant charmonium observations. Later, inspired by the Cornell model, different versions of potential models were developed by different groups \cite{Krasemann:1979ir, Stanley:1980zm, Godfrey:1985xj, Radford:2007vd, Badalian:1999fe, Barnes:2005pb}. Among these potential models, the GI model \cite{Godfrey:1985xj} was extensively applied to study meson and baryon families. In the present work, we firstly use the GI model to give the bare masses of these discussed $P-$wave charmonia, which are important input in our calculation of the unquenched quark model.

The GI model is a semirelativistic potential model which has a Hamiltonian \cite{Godfrey:1985xj}
\begin{equation}\label{Hamiltonian}
\begin{split}
H=\sqrt{\textbf{p}^2+m_1^2}+\sqrt{\textbf{p}^2+m_2^2}+\tilde{V}(\textbf{p,r}),
\end{split}
\end{equation}
where $m_1$ and $m_2$ are the masses corresponding to the quark and antiquark, and $\textbf{p}$ is the relative momentum in the center-of-mass frame. The potential between quark and antiquark is represented by $\tilde{V}(\textbf{p,r})$, which is composed of the long-range linear color confinement interaction $S(r)=br+c$ and the short-range one-gluon-exchange interaction $G(r)=-4\alpha_s(r)/(3r)$. In the nonrelativistic limit, the potential $\tilde{V}(\textbf{p,r})$ can be simplified as the familiar nonrelativistic potential. In the GI model, the relativistic correction is considered by smear transformation and the momentum-dependant factors. With the smear transformation, the above interaction can be smeared as
\begin{equation}\label{smear}
\tilde{G}(r)(\tilde{S}(r))=\int d^3r^\prime \rho_{ij}(\textbf{r}-\textbf{r}^\prime) G(r^\prime)(S(r^\prime)),
\end{equation}
where $\rho_{ij}(\textbf{r}-\textbf{r}^\prime)$ is the smearing function with the detail form, i.e.
\begin{eqnarray}
\rho_{ij}(\textbf{r}-\textbf{r}^\prime)=\frac{\sigma^3_{ij}}{\pi^{\frac{3}{2}}}e^{-\sigma_{ij}^2 (\textbf{r}-\textbf{r}^\prime)^2}
\end{eqnarray}
with
\begin{eqnarray}
\sigma_{ij}^2=\sigma_0^2\left[\frac{1}{2}+\frac{1}{2}\left[\frac{4m_i m_j}{(m_i+m_j)^2}\right]^4\right]+s^2\left[\frac{2m_i m_j}{m_i+m_j}\right],
\end{eqnarray}
Here, $m_i (m_j)$ denotes the quark mass, while $\sigma_0$ and $s$ are the universal parameters \cite{Godfrey:1985xj}. Additionally, $\pi$ is the circular constant.

In the center-of-mass system, the relativistic potential depends on the momenta of the interacting quark and antiquark. Hence, the semirelativistic correction with the momentum dependance is introduced as \cite{Godfrey:1985xj}
\begin{eqnarray}
\tilde{V}_i(r) \rightarrow\left(\frac{m_cm_{\bar{c}}}{E_c E_{\bar{c}}}\right)^{1 / 2+\epsilon_i}\tilde{V}_i(r)\left(\frac{m_cm_{\bar{c}}}{E_c E_{\bar{c}}}\right)^{1 / 2+\epsilon_i}.
\end{eqnarray}
In this equation, we define $E_c=(p^2+m_c^2)^{1/2}$ and $E_{\bar{c}}=(p^2+m_{\bar{c}}^2)^{1/2}$, and the parameter $\epsilon_i$ refers to different types of interactions. In Ref.~\cite{Godfrey:1985xj}, the details of the GI model can be found.

{To fix the parameters in the GI model, we should reproduce the mass spectrum of charmonia with the well established low lying states ($\eta_c$, $\eta_c(2S)$ $J/\psi$, $\psi(3686)$, $\chi_{c0}(1P)$, $\chi_{c1}(1P)$, $\chi_{c2}(1P)$, and $h_c(1P)$) as employed in our former work~\cite{Duan:2020tsx}. Here, the reproduced mass of the $1^1S_0$, $1^3S_1$, $2^1S_0$, $2^3S_1$, $3^3S_1$, $1^1P_1$, $1^3P_0$, $1^3P_1$, $1^3P_2$, $1^3D_1$, $1^3D_2$, $1^3D_3$, and $2^3D_1$ states of charmonium family are $2.996$, $3.098$, $3.634$, $3.676$, $4.090$, $3.513$, $3.417$, $3.500$, $3.549$, $3.805$, $3.828$, $3.841$, and $4.172$, respectively, where all values are in unit of GeV.
With the parameters as input, the low-lying charmonia can be reproduced well.} In Table~\ref{Tab2}, the bare masses of these discussed $P-$wave charmonia are collected.

\begin{table}[htbp]
\caption{The bare masses of these discussed $P-$wave charmonia give by the GI model. (Units: GeV)}
\label{Tab2}
\renewcommand\arraystretch{1.20}
\begin{tabular*}{86mm}{@{\extracolsep{\fill}}cccccc}
\toprule[1.0pt]
\toprule[1.0pt]
states &$1P$   &$2P$   &$3P$   &$4P$   &$5P$   \\%&$6P$   &$7P$   \\
\toprule[0.8pt]
$\chi_{c0}(nP)$&3.417&3.885&4.256&4.574&4.849\\%&5.122&5.367\\
$\chi_{c1}(nP)$&3.500&3.936&4.294&4.606&4.887\\%&5.146&5.388\\
$\chi_{c2}(nP)$&3.549&3.974&4.327&4.635&4.914\\%&5.171&5.411\\
\bottomrule[1.0pt]
\bottomrule[1.0pt]
\end{tabular*}
\end{table}

\begin{center}
\begin{figure}[htbp]
\includegraphics[width=7.8cm,keepaspectratio]{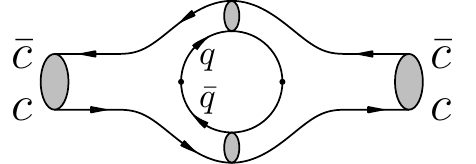}
\caption{The self-energy hadronic loops of the higher $P-$wave charmonia. Here, $q=u,d,s$ and the hadronic loops are composed of charmed and charmed strange mesons.}
\label{ccloop}
\end{figure}
\end{center}

Under the unquenched picture, the coupled-channel effect due to the interaction between the bare state and intermediate mesonic loops should be emphasized, which shown in Fig.~\ref{ccloop}. Through the intermediate meson loops, the bare state is dressed, and the mass of the bare state is shifted to a physical mass.

To quantify the self-energy correction, the coupled-channel equation is constructed as \cite{Heikkila:1983wd, Pennington:2007xr, Zhou:2011sp, Duan:2020tsx}
\begin{equation}\label{inversepropagator}
\textbf{P}^{-1}(s)\equiv M_{\rm bare}^2-s+\Pi(s)=0,
\end{equation}
where $M_{\rm bare}$ and $\Pi(s)$ are the bare mass and self-energy function, respectively. $s$ represents a pole on the complex energy plane. Since the condition $\textbf{P}^{-1}(s)=0$ is fulfilled, the coupled-channel result is determined as $s=(M_{\rm phy}-i\Gamma/2)^2$. In the result, $M_{\rm phy}$ and $\Gamma$ are the physical mass and decay width corresponding to a physical state, which can be compared with the experimental results directly. The self-energy function $\Pi(s)$ is the summation of $\Pi_n(s)$, where the subscript denotes the $n-$th hadronic channel coupled with the bare $c\bar{c}$ state. In the charmonia system, the narrow width approximation $s\approx M_{\rm phy}^2-{\rm i}M_{\rm phy}\Gamma$ works well, which was employed in Eq.~(\ref{inversepropagator}). Then, the real part and the imaginary parts of Eq.~(\ref{inversepropagator}) are separated, {\it i.e.},
\begin{equation}\label{Narrow}
\begin{split}
M_{\rm phy}^2=&M_{\rm bare}^2+{\rm Re}\Pi(M_{\rm phy}^2),\\
\Gamma=&-\frac{{\rm Im}\Pi(M_{\rm phy}^2)}{M_{\rm phy}},
\end{split}
\end{equation}
where $M_{\rm phy}$ and $\Gamma$ can be calculated directly. In Eq.~(\ref{Narrow}), with the first equation, the physical mass can be determined, then the decay width is also obtained with the second equation. We find that it is an easy way to get the coupled-channel result with Eq.~(\ref{Narrow}), which is a crucial step when investigating these higher $P-$wave charmonia.

With the dispersion relation, the imaginary part and real part of the self-energy function are linked by the integral, {\it i.e.},
\begin{equation}\label{RePi}
{\rm Re}\Pi_n(s)=\frac{1}{\pi}\mathcal{P}\int^{\infty} _{s_{{\rm th},n}} {\rm d}z\frac{{\rm Im}\Pi_n(z)}{z-s}.
\end{equation}
In this equation, $\mathcal{P}$ represents the principal value of an integral, and $s_{{\rm th},n}$ is the threshold of the $n-$th channel. Since the optical theorem is employed in the realistic calculation, we should take all possible intermediate channels into our calculation, which is obviously an impossible task. To solve the problem, the once subtracted dispersion relation was introduced by Pennington {\it et al.} in Ref.~\cite{Pennington:2007xr}. In the present work, we also use the once subtracted self energy function
\begin{equation}\label{subPi}
{\rm Re}\Pi_n(s)=\frac{s-s_0}{\pi}\mathcal{P}\int^{\infty} _{s_{{\rm th},n}} {\rm d}z\frac{{\rm Im}\Pi_n(z)}{(z-s)(z-s_0)},
\end{equation}
where $s_0$ represents the subtraction point. Generally, a ground state which is much lower than the threshold of the first OZI-allowed channel is chosen as a subtraction point. For the discussed charmonia family, the $J/\psi$ particle is chosen as subtraction point $\sqrt{s_0}=3.097~{\rm GeV}$.

With the method of once subtraction, we only consider the hadronic channels with threshold lower than the bare mass of the discussed state when performing the calculation of the self-energy function. Thus, we just need to consider the self-energy function with a limited number of intermediate hadronic loops, by which the coupled-channel effects become calculable.

Before calculating the real part of self-energy function, its imaginary part must be determined first, which is written as~\cite{Barnes:2007xu}
\begin{equation}\label{ImPi}
{\rm Im}\Pi_n(M^2)=-2\pi PE_BE_C \left|M^{LS}(P)\right|^2.
\end{equation}
The subscript $B$ and $C$ represent the two mesons involving in intermediate meson loops. $P$ is the momentum of meson $B$ in center-of-mass frame. This momentum can be expressed through $P=\lambda^{1/2}(M^2,M_B^2,M_C^2)/(2M)$, where $\lambda(x,y,z)=x^2+y^2+z^2-2xy-2xz-2yz$ is the K\"allen function. $E_{B(C)}$ is the energy of particle $B(C)$, which is $E_{B(C)}=\sqrt{P^2+M^2_{B(C)}}$. $M^{LS}(P)$ is the amplitude depicting the interaction between initial particle and intermediate meson loops, which is given by the quark pair creation (QPC) model~\cite{Micu:1968mk, LeYaouanc:1972vsx, Ackleh:1996yt, Blundell:1996as}. With the QPC model, the transition operator is expressed as
\begin{equation}\label{Tmatrix}
\begin{split}
\hat{T}=&-3 \gamma \sum_{m}\langle 1,m;1,-m |0,0\rangle \int {\rm d}^{3} {\bf p}_{3} {\rm d}^{3} {\bf p}_{4}\;\delta^{3}({\bf p}_{3}+{\bf p}_{4})\\
  &\times\mathcal{Y}_{1}^{m}\left(\frac{{\bf p}_{3}-{\bf p}_{4}}{2}\right)\chi_{1-m}^{34} \phi_0^{34} \omega_0^{34} b_{3}^{\dagger}({\bf p}_{3})d_{4}^{\dagger}({\bf p}_{4}),
\end{split}
\end{equation}
where $\phi_0^{34}$, $\chi^{34}$, $\omega_0^{34}$, and $\mathcal{Y}_{1}^{m}$ are flavor, spin, color, and orbital wave functions of the created quark pair, respectively. ${\bf P}_{3(4)}$, $b_3^\dag$ and $d_4^\dag$ are the momenta and creation operators of the quark and antiquark, which are created from the vacuum. The $\gamma$ in the equation depicts the strength of a quark-antiquark pair created from the vacuum.

With the transition operator, the amplitude $M^{LS}(P)$ is expressed as
\begin{equation}
\begin{split}
M^{LS}_{A\to BC}(P)&=\langle BC, LS|\hat{T}|A\rangle,\\
\end{split}
\end{equation}
where $L$ and $S$ are the relative orbital angular momentum and spin between $BC$, respectively. To calculate the amplitude quantitatively, the masses and wave functions of the initial and final states should be determined, which will be represented in the next section.

\subsection{The $3P$ charmonia}
\begin{center}
\begin{figure}[htbp]
\includegraphics[width=8.6cm,keepaspectratio]{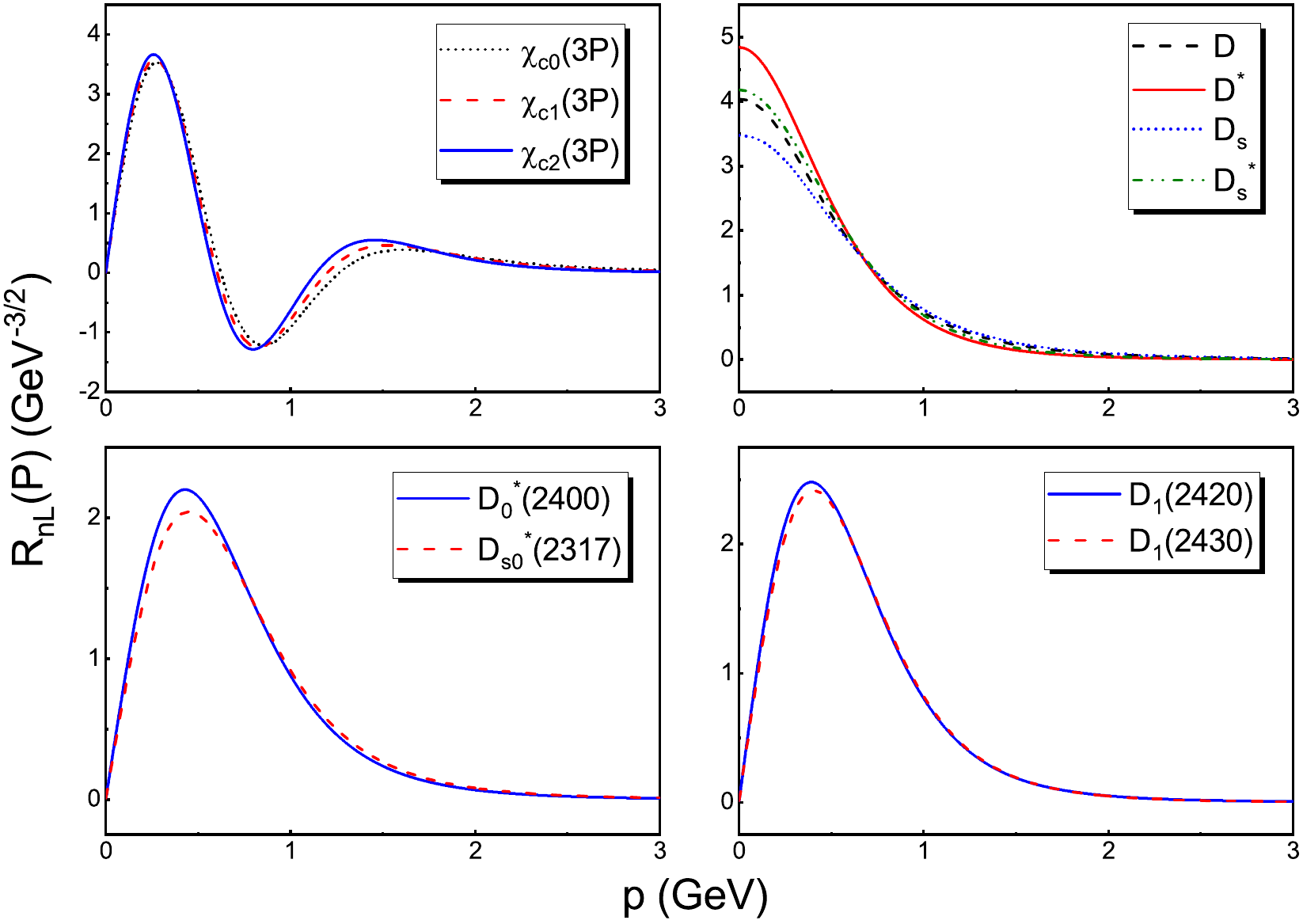}
\caption{The adopted the spatial wave functions of $\chi_{cJ}(3P)$, $D_{(s)}^{(*)}$, $D_0^*(2400)$, $D_{s0}^*(2317)$, $D_1(2420)$ and $D_1(2430)$ states in the momentum space.}
\label{3Pwf}
\end{figure}
\end{center}

With the quenched quark model, the bare masses of $3P$ charmonia are completely determined and collected in the Table~\ref{Tab2}. The physical masses of these discussed $3P$ charmonia are calculated by the coupled-channel equation, where the self-energy corrections from the intermediate hadronic loop are introduced. To quantify the self-energy function, we should take the spatial wave function as input, which can be obtained by solving the Schr\"odinger equation with the GI Hamiltonian. In Fig.~\ref{3Pwf}, the spatial wave functions of these bare $3P$ charmonia and the employed charmed mesons are shown.

Besides the spatial wave functions, the channels employed in the $3P$ charmonium calculation also need to be clarified. Here, more open-charm channels are involved in the calculation of the $3P$ charmonia, since their bare masses of $3P$ charmonia are around 4.2 GeV. In the once subtracted scheme, the contribution of the open-charm channels with mass thresholds lower than the mass of a bare state need to be summed in the self-energy function. More specifically, $D\bar{D}$, $D^*\bar{D}^*$, $D_s^+D_s^-$, and $D_s^{*+}D_s^{*-}$ channels are included in our calculation for $\chi_{c0}(3P)$. The $D\bar{D}^*$, $D^*\bar{D}^*$, $D_0^*(2400)\bar{D}$, $D_1(2420)\bar{D}$, $D_1(2430)\bar{D}$, $D_s^+D_s^{*-}$, $D_s^{*+}D_s^{*-}$, and $D_{s0}^{*+}(2317)D_s^-$ channels are considered in the coupled channel calculation of $\chi_{c1}(3P)$. For $\chi_{c2}(3P)$, the involved coupled channels are $D\bar{D}$, $D\bar{D}^*$, $D^*\bar{D}^*$, $D_1(2420)\bar{D}$, $D_1(2430)\bar{D}$, $D_s^+D_s^-$, $D_s^+D_s^{*-}$ and $D_s^{*+}D_s^{*-}$. We list the experimental masses of these employed $1S$ and $1P$ charmed and charmed-strange mesons in Table~\ref{Dmeson1}.

\begin{table}[htbp]
\caption{As input parameter, the masses of these involved charmed and charmed-strange mesons employed in the $3P$ charmonium calculation are taken from PDG~\cite{ParticleDataGroup:2020ssz}.}
\label{Dmeson1}
\renewcommand\arraystretch{1.20}
\begin{tabular*}{86mm}{@{\extracolsep{\fill}}ccccc}
\toprule[1.5pt]
\toprule[1pt]
State     &$D$        &$D^*$      &$D_s$        &$D_s^*$      \\
Mass (GeV)&1.867      &2.009      &1.968        &2.112        \\
\midrule[0.75pt]
State&$D_{0}^*(2400)$&$D_1(2420)$&$D_1(2430)$&$D_2^*(2460)$\\
Mass (GeV)&2.325          &2.422      &2.427      &2.463        \\
\midrule[0.75pt]
State&$D_{s0}^*(2317)$&$D_{s1}(2460)$&$D_{s1}(2536)$&$D_{s2}(2573)$\\
Mass (GeV)&2.317           &2.460         &2.535         &2.569         \\
\bottomrule[1pt]
\bottomrule[1.5pt]
\end{tabular*}
\end{table}

In the calculation of $3P$ charmonia, the mixing scheme of $D_1(2420)$ and $D_1(2430)$ should be considered. In the heavy quark symmetry, $D_1(2420)$ and $D_1(2430)$ are mixture bewteen the $D(1^3P_1)$ and $D(1^1P_1)$ states which satisfy the relation
\begin{equation}
\begin{split}
\begin{pmatrix} |D(2430)\rangle \\ |D(2420)\rangle\end{pmatrix}=\begin{pmatrix} {\cos} \theta & {\sin} \theta\\ {-\sin} \theta& {\cos} \theta \end{pmatrix}\begin{pmatrix} |D(1^1P_1)\rangle \\ |D(1^3P_1)\rangle \end{pmatrix} \\
\end{split},
\end{equation}
where the mixing angle was fixed to be $\theta=-54.7^\circ$ in the heavy quark limit \cite{Song:2015nia}.

Now the only unknown parameter is the coupling constant $\gamma$. Through the decay width of $\psi(3770)$ and $\psi(4160)$, the $\gamma$ value is determined as $\gamma$=0.4 \cite{Duan:2020tsx}. With the above preparation, we may predict the spectroscopy behavior of the discussed $3P$ charmonia.

\begin{center}
\begin{figure}[htbp]
\includegraphics[width=8.6cm,keepaspectratio]{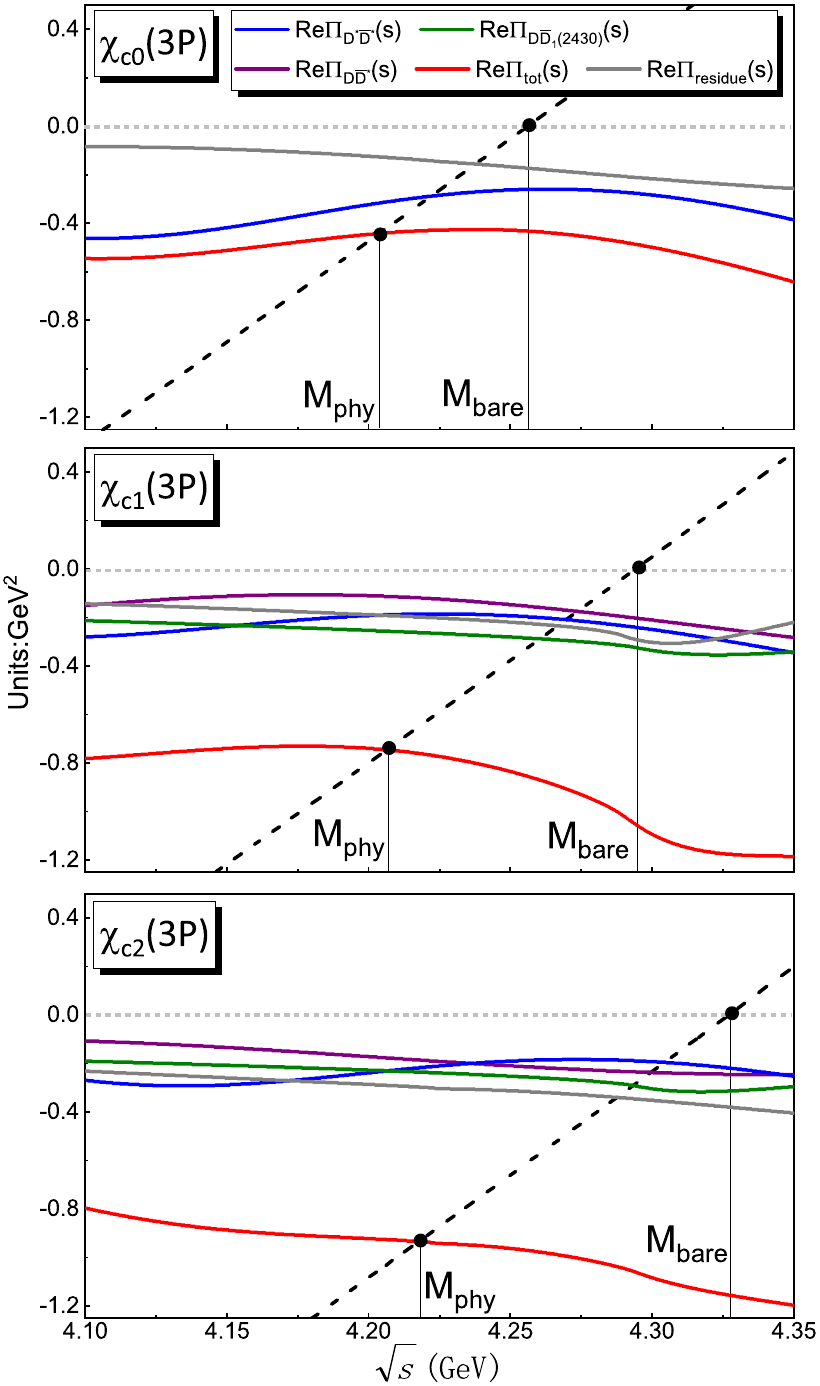}
\caption{The real part of self energy functions of the $\chi_{cJ}(3P)$ states are shown. To outstand the prominent contributions, only the major channels are plotted individually, and the contribution of the remaining channels are combined in the ${\rm Re}\Pi_{\rm residue}(s)$ part with the gray solid line. Here, the black dashed line represents the function $s-M_{\rm bare}^2$ while the red solid line is the total self-energy function. The blue, green, and purple solid lines denote the self-energy functions of the $D^*\bar{D}^*$, $D\bar{D}_1(2430)$, and $D\bar{D}^*$ channels, which provide the prominent self-energy contribution to the $3P$ charmonia.}
\label{3Pre}
\end{figure}
\end{center}

The self-energy function ${\rm Re}\Pi(s)$ and the corresponding function $s-M_{\rm bare}^2$ dependent on $s$ for each $3P$ charmonium are illustrated in Fig.~\ref{3Pre}.  The physical mass is found by the abscissa of the intersection at the cross of red solid line and the dashed line, which is the solution of the real part of coupled-channel equation in Eq.~(\ref{Narrow}).

When making comparison of the coupled-channel corrections of $\chi_{c0}(3P)$, $\chi_{c1}(3P)$, and $\chi_{c2}(3P)$ shown in the Fig.~\ref{3Pre}, we find that their mass shift between the bare mass and the physical mass are obvious and in the same order of magnitude. After including the coupled-channel correction, their physical masses still satisfy the relation $M_{\chi_{c2}(3P)}>M_{\chi_{c1}(3P)}>M_{\chi_{c0}(3P)}$. In the top diagram of Fig.~\ref{3Pre}, the comparison of the total self-energy function ${\rm Re}\Pi(s)$ (red solid line) and the self-energy function of the $D^*\bar{D}^*$ channel of $\chi_{c0}(3P)$ (blue solid line) are given, which show that the mass shift of $\chi_{c0}(3P)$ is dominated by the contribution of the $D^*\bar{D}^*$ channel. In the middle and bottom diagrams in Fig.~\ref{3Pre}, the self-energy functions of $\chi_{c1}(3P)$ and $\chi_{c2}(3P)$ are also illustrated. For the discussed $\chi_{c1}(3P)$ and $\chi_{c2}(3P)$ states, not only the $D^*\bar{D}^*$ channel provides an obvious contribution to the self-energy correction, but also the $D\bar{D}^*$ and $D\bar{D}_1(2430)$ channels also provide a large contribution. In  Table~\ref{3Pmass}, the masses and decay widths of $3P$ charmonia are listed here, where their partial decay widths and corresponding branching ratios are shown in the Table~\ref{3Pdecay}.

\begin{table}
\caption{The obtained physical masses of $\chi_{cJ}(3P)$ are listed. The bare mass, decay width and $\delta M=M_{\rm phy}-M_{\rm bare}$ are also shown here.}
\label{3Pmass}
\renewcommand\arraystretch{1.20}
\begin{tabular*}{86mm}{@{\extracolsep{\fill}}ccccc}
\toprule[1.5pt]
\toprule[1pt]
state&$M_{\rm bare}$ (GeV)&$M_{\rm phy}$ (GeV)&$\delta M$ (MeV)&$\Gamma_{tot}$ (MeV)\\
\midrule[0.75pt]
$\chi_{c0}(3P)$&4.256&4.204&-52&72\\
$\chi_{c1}(3P)$&4.294&4.206&-88&48\\
$\chi_{c2}(3P)$&4.327&4.218&-109&50\\
\bottomrule[1pt]
\bottomrule[1.5pt]
\end{tabular*}
\end{table}

\begin{table}
\caption{The partial decay widths and corresponding branching ratios are shown. The sign $\times$ means that the corresponding channel is included in the self-energy function, but their decay process is kinematically forbidden. The zero decay width listed in this table represents the corresponding decay width far smaller than 1 MeV, which can be ignored in our discussion.}
\label{3Pdecay}
\renewcommand\arraystretch{1.20}
\begin{tabular*}{86mm}{@{\extracolsep{\fill}}ccccccc}
\toprule[1.5pt]
\toprule[1pt]
\multirow{2}{*}{channels}&\multicolumn{2}{c}{$\chi_{c0}(3P)$}&\multicolumn{2}{c}{$\chi_{c1}(3P)$}&\multicolumn{2}{c}{$\chi_{c2}(3P)$}\\
\Xcline{2-3}{0.75pt}
\Xcline{4-5}{0.75pt}
\Xcline{6-7}{0.75pt}
&$\Gamma$ (MeV)&$\mathcal{B}$ (\%)&$\Gamma$ (MeV)&$\mathcal{B}$ (\%)&$\Gamma$ (MeV)&$\mathcal{B}$ (\%)\\
\midrule[1pt]
$D\bar{D}$               &2       &3       &           &          &7         &14  \\
$D\bar{D}^*$             &        &        &20         &42        &3         &6   \\
$D^*\bar{D}^*$           &67      &93      &26         &54        &39        &78  \\
$D\bar{D}_0^*(2400)$     &        &        &$\sim0$    &$\sim0$   &          &    \\
$D\bar{D}_1(2420)$       &        &        &$\times$   &$\times$  &$\times$  &$\times$\\
$D\bar{D}_1(2430)$       &        &        &$\times$   &$\times$  &$\times$  &$\times$\\
$D_s^+D_s^-$             &3       &4       &           &          &$\sim$0   &$\sim$0   \\
$D_s^+D_s^{*-}$          &        &        &2          &4         &1         &2   \\
$D_s^{*+}D_s^{*-}$       &$\times$&$\times$&$\times$   &$\times$  &$\times$  &$\times$\\
$D_s^+D_{s0}^{*-}(2317)$ &        &        &$\times$   &$\times$  &          &    \\
\midrule[1pt]
Total&72&100&48&100&50&100\\
\bottomrule[1pt]
\bottomrule[1.5pt]
\end{tabular*}
\end{table}

Our calculation shows that the physical mass of $\chi_{c0}(3P)$ is 4.204 GeV. Its mass gap between the bare and physical mass is $M_{\rm phy}-M_{\rm bare}=$-52 MeV. And the total decay width is determined as $\Gamma_{\chi_{c0}(3P)}=72$ MeV. In these allowed decay modes of $\chi_{c0}(3P)$, the $D^*\bar{D}^*$ channel has the largest partial decay width and branching ratio, which are calculated to be $\Gamma^{D^*\bar{D}^*}_{\chi_{c0}(3P)}=66$ MeV and $\mathcal{B}(\chi_{c0}(3P)\to D^*\bar{D}^*)=$93\%, respectively, which makes us understand why the $D^*\bar{D}^*$ channel plays crucial role in the coupled-channel analysis to the $\chi_{c0}(3P)$ state as shown in Fig.~\ref{3Pre}.
The physical mass of the $\chi_{c1}(3P)$ state is $M_{\rm phy}=4.206$ GeV, which is close to the physical mass of the $\chi_{c0}(3P)$ state, but the mass shift of $\chi_{c1}(3P)$ is $M_{\rm phy}-M_{\rm bare}=$-88 MeV. Its total decay width is 48 MeV, where the partial decay widths of the $D^*\bar{D}^*$ and $D\bar{D}^*$ channel are 26 MeV and 20 MeV, respectively. With the partial decay widths, the corresponding branching ratios of $D^*\bar{D}^*$ and $D\bar{D}^*$ channel can be obtained to be 54\% and 42\%, respectively. In addition, the partial decay widths of the $D\bar{D}_0(2400)$ and $D_s^+D_s^{*-}$ channels are 0.1 MeV and 2 MeV, respectively.
The physical mass of the remaining state $\chi_{c2}(3P)$ is $M_{\rm phy}=$4.218 MeV, which has 109 MeV mass shift compared with bare mass $M_{\rm bare}=4.327$ GeV.
The total decay width of $\chi_{c2}(3P)$ is $\Gamma_{\chi_{c2}(3P)}=$50 MeV, which is composed of the partial decay widths $\Gamma_{\chi_{c2}(3P)}^{D^*\bar{D}^*}=$39 MeV, $\Gamma_{\chi_{c2}(3P)}^{D\bar{D}}=$7.0 MeV, $\Gamma_{\chi_{c2}(3P)}^{D\bar{D}^*}=$3 MeV, $\Gamma_{\chi_{c2}(3P)}^{D_s^+D_s^-}=$0.1 MeV and $\Gamma_{\chi_{c2}(3P)}^{D_s^+D_s^{*-}}=$1 MeV. The branching ratios of the $D^*\bar{D}^*$ and $D\bar{D}$ decay channels are 78\% and 14\%, respectively, where the main decay channel for $\chi_{c2}(3P)$ is $D^*\bar{D}^*$.

In summary, as demonstrated above, the physical masses of $\chi_{c0}(3P)$, $\chi_{c1}(3P)$ and $\chi_{c2}(3P)$ are close to each other, and these $3P$ charmonia are not narrow states.
How to distinguish these $3P$ states becomes a challenge in
experimental analysis if these $3P$ states appear in the same decay channel.

At present, the experimental data of the $\gamma\gamma\to D\bar{D}$ \cite{Uehara:2005qd, BaBar:2010jfn} and $B\to KD\bar{D}$ \cite{Aaij:2020ypa, LHCb:2020bls} processes reveal the existence of the $2P$ charmonia around 3.9 GeV.
Thus, searching for the $3P$ charmonia via $\gamma\gamma$ fusion process and $B$ decay
is suggested, {\it i.e.,} we encourage our experimental colleague to measure $\gamma\gamma\to D^{(*)}\bar{D}^{(*)}$ and $B\to KD^{(*)}\bar{D}^{(*)}$ and analyze the corresponding $D^{(*)}\bar{D}^{(*)}$ invariant mass spectrum.

Of course, hunting for these $3P$ charmonia via their hidden-charm decay channels from $\gamma\gamma$ fusion process and $B$ decay is an interesting issue.
Until now, experiment has reported two charmoniumlike states
$Y(4140)$ and $Y(4274)$ by measuring the $B\to KJ/\psi\phi$ process \cite{LHCb:2016axx, CDF:2011pep}, which have mass close 4.2 GeV.
{Meanwhile, the $Y(4140)$ and $Y(4274)$ have spin-parity quantum number with $J^{PC}=1^{++}$, and the possibility of determining $Y(4140)$ as a $\chi_{c1}(3P)$ state is proposed by the narrow decay width of $\chi_{c1}(3P)$ state through an explicit calculation with unquenched quark model and QPC model in Ref.~\cite{Hao:2019fjg}. The similar opinion is also indicated in Ref.~\cite{Chen:2016iua}. By fitting the data in the $B\to K\chi_{c1}(1P)\pi\pi$ process, a narrow state around $m=4144.5$ GeV is found, which is assumed as a same state with $Y(4140)$ and suggested as a candidate of $\chi_{c1}(3P)$}.
However, since $Y(4274)$
%favors the spin-parity quantum number $J^{PC}=1^{++}$ and
has full width $\Gamma=49\pm12$ MeV \cite{ParticleDataGroup:2020ssz} consistent with our theoretical result, it is also possible to assign $Y(4274)$ as the $\chi_{c1}(3P)$ state. This conclusion was also made in Refs.~\cite{Ferretti:2020civ, Giron:2020qpb}. For enhancing this conclusion, further theoretical and experimental studies around these $3P$ charmonia are needed in near future.

\subsection{Higher $P$-wave states}

In the above section, we present the results of the $3P$ charmonia via an unquenched quark model. In the following, we may continue to discuss higher $P$-wave charmonium states involved in these $4P$ and $5P$ states.

First, we focus on the $4P$ charmonia.
With the once subtracted scheme, there is 15 intermediate channels should be included in the self-energy function of the $\chi_{c0}(4P)$ states, which are the $D\bar{D}$, $D^*\bar{D}^*$, $D\bar{D}_1(2420)$, $D\bar{D}_1(2430)$, $D^*\bar{D}^*_0(2400)$, $D^*\bar{D}_1(2420)$, $D^*\bar{D}_1(2430)$, $D\bar{D}(2^1S_0)$, $D^*\bar{D}^*_2(2460)$, $D_s^+D_s^-$, $D_s^{*+}D_s^{*-}$, $D_s^+D^-_{s1}(2460)$, $D_s^{*+}D_{s0}^{*-}(2317)$, $D_s^{*+}D_{s1}^-(2536)$, $D_s^{*+}D_{s1}^-(2460)$ channels.
For $\chi_{c1}(4P)$ and $\chi_{c2}(4P)$, there are 20 and 22 intermediate hadronic loops contained in ${\rm Re}\Pi(s)$ function. Thus, the whole calculation of the $4P$ states becomes more complicated compared with former study of the $3P$ state. In Table \ref{Dmeson2}, the involved charmed and charmed-strange mesons are summarized with the corresponding masses as input.

\begin{table}[htbp]
\caption{The adopted masses of these involved $1D$, $2S$, and $2P$ charmed and charmed strange mesons are listed, which are taken from Ref.~\cite{Godfrey:1985xj}}
\label{Dmeson2}
\renewcommand\arraystretch{1.20}
\begin{tabular*}{86mm}{@{\extracolsep{\fill}}cccccc}
\toprule[1.5pt]
\toprule[1pt]
State     &$D(2^1S_0)$        &$D(2^3S_1)$      &$D(2^3P_0)$        &$D(2^3P_1)$     &$D(2^1P_1)$ \\
Mass (GeV)&2.583              &2.645            &2.932              &2.952           &2.933\\
   State       &$D(2^3P_2)$        &$D(1^1D_2)$      &$D(1^3D_1)$        &$D(1^3D_2)$     &$D(1^3D_3)$ \\
       Mass (GeV)   &2.957              &2.827            &2.816              &2.834           &2.833\\
\midrule[0.75pt]
   State       &$D_s(2^1S_0)$      &$D_s(2^3S_1)$    &$D_s(2^3P_0)$      &$D_s(2^3P_1)$   &$D_s(2^1P_1)$\\
    Mass (GeV)      &2.675              &2.735            &3.005              &3.033           &3.024\\
      State    &$D_s(2^3P_2)$      &$D_s(1^1D_2)$    &$D_s(1^3D_1)$      &$D_s(1^3D_2)$   &$D_s(1^3D_3)$\\
    Mass (GeV)      &3.049              &2.910            &2.898              &2.915           &2.916\\
\bottomrule[1pt]
\bottomrule[1.5pt]
\end{tabular*}
\end{table}

With the above preparations, the coupled-channel results for the $4P$ charmonia are given in Table~\ref{4Pmass}. And the partial decay width of the discussed $4P$ charmonia are shown in Table~\ref{4Pdecay}. Our results show that the coupled-channel correction to the masses of $4P$ charmonia is about 200 MeV, which makes their physical masses become $M_{\rm phy}(\chi_{c0}(4P))=$4.358 GeV, $M_{\rm phy}(\chi_{c1}(4P))=$4.378 GeV, and $M_{\rm phy}(\chi_{c2}(4P))=$4.397 GeV. In addition, we also obtain the total decay widths of these $4P$ charmonia as shown in Table~\ref{4Pdecay}, where the $D^*\bar{D}^*$ channel still has main contribution to the full widths of $\chi_{cJ}(4P)$ states.

For the $5P$ charmonia, more channels should be considered, where $\chi_{c0}(5P)$, $\chi_{c1}(5P)$, and $\chi_{c2}(5P)$ have 33, 53, and 56 intermediate channels, respectively when performing the coupled-channel analysis. In Table \ref{4Pmass}, we predict the masses of $\chi_{c0}(5P)$, $\chi_{c1}(5P)$, and $\chi_{c2}(5P)$.

\iffalse
\begin{table}[htbp]
\caption{The physical masses and corresponding decay widths of $4P$, $5P$ states with a coupled-channel analysis are listed here.}
\label{4Pmass}
\renewcommand\arraystretch{1.20}
\begin{tabular*}{86mm}{@{\extracolsep{\fill}}cll}
\toprule[1.5pt]
\toprule[1pt]
                                  &$4P$      &$5P$      \\  %&$6P$         &$7P$   \\
\toprule[0.75pt]
$M_{\rm phy}(\chi_{c0}(nP))$ (GeV)&4.358   &4.504     \\%&4.682      &4.769\\
$\Gamma$ (MeV)                    &43      &17       \\% &15         &4    \\

\toprule[0.75pt]

$M_{\rm phy}(\chi_{c1}(nP))$ (GeV)&4.378   &4.511    \\% &4.696      &4.770\\
$\Gamma$ (MeV)                    &41      &19       \\% &16         &4    \\

\toprule[0.75pt]

$M_{\rm phy}(\chi_{c2}(nP))$ (GeV)&4.397   &4.524  \\%   &4.703      &4.778\\
$\Gamma$ (MeV)                    &27      &15       \\% &10         &2    \\
\bottomrule[1pt]
\bottomrule[1.5pt]
\end{tabular*}
\end{table}
\fi

\begin{table}
\caption{The physical masses and corresponding decay widths of $4P$, $5P$ states with a coupled-channel analysis are listed here.}
\label{4Pmass}
\renewcommand\arraystretch{1.20}
\begin{tabular*}{86mm}{@{\extracolsep{\fill}}ccccc}
\toprule[1.5pt]
\toprule[1pt]
state&$M_{\rm bare}$ (GeV)&$M_{\rm phy}$ (GeV)&$\delta M$ (MeV)&$\Gamma_{tot}$ (MeV)\\
\midrule[0.75pt]
$\chi_{c0}(4P)$&4.574&4.358&-216&43\\
$\chi_{c1}(4P)$&4.606&4.378&-228&41\\
$\chi_{c2}(4P)$&4.635&4.397&-238&27\\
\midrule[0.75pt]
$\chi_{c0}(5P)$&4.849&4.504&-345&17\\
$\chi_{c1}(5P)$&4.887&4.511&-376&19\\
$\chi_{c2}(5P)$&4.914&4.524&-390&15\\
\bottomrule[1pt]
\bottomrule[1.5pt]
\end{tabular*}
\end{table}

\begin{table}[htbp]
\caption{The calculated partial decay widths and corresponding branching ratios of the $\chi_{cJ}(4P)$ charmonia are listed. Here, we only show the channels which have decay widths larger than 1 MeV. }
\label{4Pdecay}
\renewcommand\arraystretch{1.20}
\begin{tabular*}{86mm}{@{\extracolsep{\fill}}cccccc}
\toprule[1.5pt]
\toprule[1pt]

 $\chi_{c0}(4P)$     & $D\bar{D}$       & $D^*\bar{D}^*$      & $D_1(2420)\bar{D}$     & $D_0(2400)\bar{D}^*$      & $D_s^+D_s^-$  \\
 $\Gamma$ (MeV)      & 7                & 22                  & 10                     & 3                         & 1             \\
 $\mathcal{B}$ (\%)  & 16               & 51                  & 23                     & 7                         & 2             \\

 \midrule[0.75pt]
 $\chi_{c1}(4P)$     & $D\bar{D}^*$     & $D^*\bar{D}^*$      & $D_2^*(2460)\bar{D}$   & $D_0(2400)\bar{D}^*$      &               \\
 $\Gamma$ (MeV)      & 15               & 12                  & 10                     & 3                         &               \\
 $\mathcal{B}$ (\%)  & 37               & 29                  & 24                     & 7                         &               \\

 \midrule[0.75pt]
 $\chi_{c2}(4P)$     & $D\bar{D}$       & $D^*\bar{D}^*$      & $D_1(2420)\bar{D}$     & $D_1(2430)\bar{D}$        & $D_2(2460)\bar{D}$ \\
 $\Gamma$ (MeV)      & 2                & 17                  & 3                      & 2                         & 3                  \\
 $\mathcal{B}$ (\%)  & 7                & 63                  & 11                     & 7                         & 11\\

\bottomrule[1pt]
\bottomrule[1.5pt]
\end{tabular*}
\end{table}

\begin{center}
\begin{figure}[htbp]
\includegraphics[width=8.6cm,keepaspectratio]{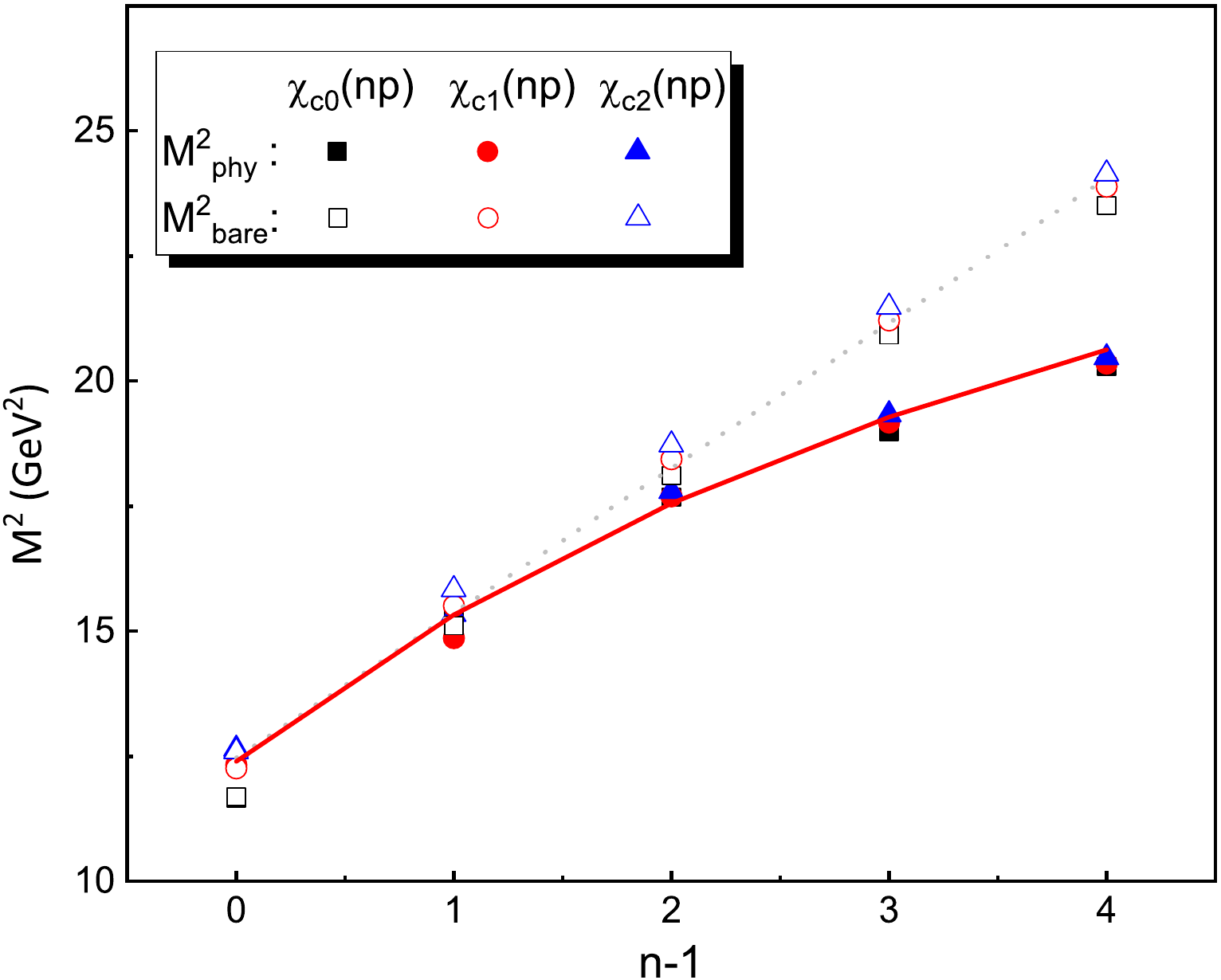}
\caption{The comparison of the obtained physical and bare masses for the discussed $P$-wave charmonia are given. Here, the red solid curve is the result of the modified GI model when taking $\mu=0.13$ GeV.}
\label{SQmass}
\end{figure}
\end{center}

Now, we obtained mass spectrum of higher $P-$wave charmonia with an unquenched quark model. In Fig.~\ref{SQmass}, we make comparison of the bare masses and physical masses of higher $P-$wave charmonia with different radial quantum numbers. We find that  the bare masses are linear relation with increasing radial quantum number $n$. When the coupled-channel effect included, such linear relation was violated, which can be understood by the modified GI model.

{As unquenched potential model, }
the modified GI model was employed in Refs. \cite{Wang:2019mhs, Wang:2020prx}, which was applied to study higher vector charmonia.
Here, the linear confinement interaction in the GI model (see Sec.~\ref{subsec21}) is replaced by the screened confinement interaction Ref.~\cite{Li:2009ad}, {\it i.e.,}
\begin{equation}
\begin{split}
S(r)=br+c\longrightarrow S^{scr}(r)=\frac{b(1-e^{-\mu r})}{\mu}+c,\\
\end{split}
\end{equation}
where an additional parameter $\mu$ is appeared in the screened potential. With selecting the parameter $\mu$, the strength of the screened confinement interaction can be controlled. We find that the mass spectrum behavior of the obtained $P$-wave charmonia shown in Fig.~\ref{SQmass} can be mimicked by the modified GI model when $\mu=0.13$ GeV.
{By this study, we also illustrate why
the modified GI model by introducing the screened potential can achieve
a similar result as obtained by the coupled-channel model~\cite{Li:2009ad}.}

\section{Summary}\label{sec3}

After establishing $2P$ charmonia \cite{Duan:2020tsx, Liu:2009fe, Chen:2012wy, Chen:2013yxa, Aaij:2020ypa, LHCb:2020bls}, how to explore higher $P$-wave charmonium states becomes an intriguing research issue, especially, with the running of Belle II \cite{Belle-II:2018jsg}, and Run-II and Run-III at LHCb \cite{LHCb:2018roe}. Obviously, valuable hints can be given by theoretical investigation around higher $P$-wave charmonium states.

In this work, we first focus on the $3P$ charmonia by presenting their spectroscopy behavior. In fact, the studying of these $2P$ charmonia revealed the crucial role of the coupled-channel effect on mass spectrum \cite{Duan:2020tsx}. Thus, we adopt an unquenched quark model to depict the spectroscopy behavior of these $3P$ charmonia. Our result shows that these $3P$ states have mass around 4.2 GeV. Hunting for these $3P$ states via their open-charm and hidden-charm decay channels from $\gamma\gamma$ fusion and $B$ decay is suggested.

Under the same framework, we may continue to predict the spectroscopy behavior of $4P$ and $5P$ charmonia. We find that mass shift of physical mass and bare mass becomes more obvious with increasing the radial quantum number $n$. For understanding this phenomenon, we take the modified GI model to mimic the spectroscopy behavior of higher $P$-wave charmonia, {which shows that
the modified GI model and the coupled-channel model, both of which are unquenched quark model,
can get the similar result when studying meson spectroscopy. }

{When studying $XYZ$ data \cite{LHCb:2021uow, LHCb:2016axx, Belle:2007woe, Belle:2009rkh}, multiquark state explanations and some other possible effects resulting in these novel phenomena was proposed \cite{Liu:2009ei, Liu:2009qhy, Liu:2010hf, Mahajan:2009pj, Ding:2009vd, Finazzo:2011he, Wang:2011uk, Lu:2016cwr, Anwar:2018sol, Maiani:2016wlq, Chen:2016oma, Liu:2016onn}. For the coupled-channel effect, there still exist large uncertainties in the calculation. Thus, how to distinguish these contributions from that from the coupled-channel effect should be paid more attention especially when identifying $3P$ and higher $P$-wave charmonia by combing with the present $XYZ$ data.}

As emphasized by the present experimental progress, Belle II and LHCb will become the main force of finding out charmoniumlike states. In the past 18 years, the experiments have brought us big surprises. In the following years, we have reason to believe that
the study of hadron spectroscopy enters a new era. As an important part of hadron spectroscopy, we should pay more attention to the charmonium family. We hope that the present work can inspire the experimentalist's enthusiasm of exploring higher states of charmonium family.

%******************************************************************
\section*{Acknowledgements}
%******************************************************************
This work is supported by the China National Funds for Distinguished Young Scientists under Grant No. 11825503, National Key Research and Development Program of China under Contract No. 2020YFA0406400, the 111 Project under Grant No. B20063, the National Natural Science Foundation of China under Grants No. 12047501, and the Fundamental Research Funds for the Central Universities.

%**********************************************************************************************************************
%**********************************************************************************************************************

%******************************************************************
\end{document}